\documentclass[11pt]{article}
\usepackage{hyperref}
\pdfoutput=1

\begin{document}
\title{Hydrothermal waves in evaporating sessile drops (APS 2009)}
\author{D. Brutin, F. Rigollet, C. LeNiliot \\
\\ \vspace{6pt} Aix-Marseille University, IUSTI Laboratory \\ 13013 Marseille, FRANCE}

\maketitle


\begin{abstract}
This fluid dynamics video was submitted to the Gallery of Fluid Motion for the 2009 APS Division of Fluid Dynamics Meeting in Minneapolis, Minnesota. Drop evaporation is a simple phenomena but still unclear concerning the 
mechanisms of evaporation. A common agreement of the scientific community based on experimental and numerical work evidences that most of the evaporation occurs at the triple line. However, the rate of evaporation is still 
empirically predicted due to the lack of knowledge on the convection cells which develop inside the drop under evaporation. The evaporation of sessile drop is more complicated than it appears due to the coupling by conduction with 
the heating substrate, the convection and conduction inside the drop and the convection and diffusion with the vapour phase. The coupling of heat transfer in the three phases induces complicated cases to solve even for numerical 
simulations. We present recent experimental fluid dynamics videos obtained using a FLIR SC-6000 coupled with a microscopic lens of 10 $\mu$m of resolution to observe the evaporation of sessile drops in infrared wavelengths. The 
range of 3 to 5 $\mu$m is adapted to the fluids observed which are ethanol, methanol and FC-72 since they are all half-transparent to the infrared.
\end{abstract}


\section{Description}

The movie sample presented is dealing with methanol sessile drops under evaporation. Three different fluids are used \href{http://ecommons.library.cornell.edu/bitstream/1813/14131/2/METHANOL.m2v}{Methanol}, Ethanol and FC-72. The 
original movie presended here is recorded with a resolution of 640x512 pixels. This movie is presented at real time speed (25 frames/sec). To fit the web site request, the video quality have been sharply reduced. For the three 
cases, the substrate is PFTE at constant surface temperature and with a surface roughness of 400 nm and no specific direction. The surface temperature is regulated to be constant using a PID heating device coupled with heating 
cartridges. The ambient temperature for all three cases is 25$^{\circ}$C and the atmosphere is air at 1 atm.

\begin{table}[ht]
\begin{center}
\caption{Physical properties of fluids at 25$^{\circ}$C and 1 atm}
\vspace{2mm}
\label{fluids}
{\renewcommand{\arraystretch}{1.4}
\footnotesize
\begin{tabular}{cccccccc}
\hline
          & $\varrho_{L}$& Cp                   & Lv             & $\mu$  & $\sigma$    & T$_{sat}$    & Pr       \\
\hline
          & kg.m$^{-3}$  & J.kg$^{-1}$.K$^{-1}$ & kJ.kg$^{-1}$   & mPa.s  & mN.m$^{-1}$ & $^{\circ}$C       & -        \\
\hline  
Water     & 997          & 4180                 & 2449           & 0.890  & 72.7        & 100              & 6.14     \\
FC-72     & 1680         & 1100                 & 88.0           & 0.638  & 12.0        & 56.0             & 12.3    \\
Methanol  & 791          & 2531                 & 1165           & 0.560  & 22.7        & 64.7             & 6.98     \\
Ethanol   & 789          & 2845                 & 841            & 1.095  & 22.0        & 78.0             & 22.3     \\
\hline
\end{tabular}}
\end{center}
\end{table} 

Initially, the substrate is at a constant temperature. This is checked by the first image of the video. The thermal homogeneity is less than 1$^{\circ}$C at the beginning the experiment. All fluids for these thickness are 
half-transparent. The different steps of a volatile drop evaporation can be distinguished as follow:

\begin{enumerate}
\item Phase 1 - Drop warming up - The start of the experiment is defined as the initial contact of the drop posed on the substrate. The first seconds of the experiments are characterized by a transition phenomena, the drop which 
was initially at room temperature is posed on a warm surface, the drop first is heated to reach almost the substrate temperature. The first step of the phenomena is thus only driven by the fluid heat capacity.

\item Phase 2 - Drop evaporation - The heat flux transferred to the drop reach a maximum value which correspond the beginning of our evaporation investigation. The drop is under evaporation, the heat flux linearly decrease during 
a first phase which correspond to the existence of convective cell inside the drop. The flow motion inside the drop depends on the fluid physical properties such as the fluid viscosity and latent heat of vaporization.

\item Phase 3 - Film evaporation -  The last step of evaporation is characterized by the sharp decrease of the heat flux when the convection cells are vanished. No more fluid motion can be observed inside the drop. An explanation 
can be provided based on the critical thickness for thermal flow instabilities to develop. 
\end{enumerate}

Below, we present three experimental fluid dynamics videos for three different liquids with one objective: show up the complexity of drop evaporation mechanism.

\begin{enumerate}
\item In the case of \href{http://ecommons.library.cornell.edu/bitstream/1813/14131/2/METHANOL.m2v}{Methanol}, numerous hydrothermal waves can be observed at the first stage of the evaporation near the drop perimeter. In the 
center of the drop, the flow is much more unstable. The thermal flow motion inside the drop is fast, compared to the other fluids. The flow motion inside the drop stop when the drop reach 2 mm in diameter. 

\item In the case of Ethanol, it is possible to observe with a greyscale evidence the temperature isotherm which are located in the middle of the drop. Near the drop perimeter, less hydrothermal waves are observed. Then during the 
evaporation, the thermal flow change very quickly to big hydrothermal waves inside the drop. The time of evaporation is comparable to the methanol case since the heat of vaporization is comparable.

\item In the case of FC-72, the latent of vaporization is at least 10 times smaller compared to methanol and ethanol. Consequently even with a substrate temperature at 29$^{\circ}$C (slightly above the room temperature), the drop 
evaporation time is fast compared to the two other liquids. During the FC-72 evaporation, multiple internal convective cell structures can be observed which is completely different in the case of methanol and ethanol evaporation 
dynamics.
\end{enumerate}

To conclude, for fluids with even comparable latent heat of vaporization, the thermal flow motion inside the drop can be very different since the fluid viscosity is the only important difference in the fluid physical properties. 
With the knowledge of the temperature scale, it is possible to relate the temperature difference inside the drop up to 10$^{\circ}$C for methanol to the quick flow motion; where as for the ethanol drop, the maximum temperature 
difference is only 5$^{\circ}$C. For very low heat of vaporization fluid (FC-72 for example) the dynamic of evaporation is completely different with quite no more hydrothermal waves but several convection cells inside the drop. 
The FC-72 drop evaporation is closer to a FC-72 layer evaporation which is another research area of investigation.

\end{document}